\documentclass[prb,twocolumn,aps,floats,showpacs]{revtex4}
\usepackage{graphicx}%

\begin{document}
\title{ \bf Self-Trapping of Polarons in the Rashba-Pekar Model}
\author{
A.S.~Mishchenko$^{1,2}$, N.~Nagaosa$^{1,3}$, N.V.~Prokof'ev$^{4}$,
A.~Sakamoto$^3$, and B.V.~Svistunov$^2$ }
\affiliation{
$^1$Correlated Electron Research Center, AIST, Tsukuba Central 4, 
Tsukuba 305-8562, Japan
\\ $^2$Russian Research Center ``Kurchatov Institute", 123182, 
Moscow, Russia \\
$^3$Department of Applied Physics, The University of Tokyo, 7-3-1
Hongo, Bunkyo-ku, Tokyo 113, Japan \\ $^4$ Department of Physics,
University of Massachusetts, Amherst, MA 01003, USA }
\begin{abstract}
We performed quantum Monte Carlo study of the exciton-polaron
model which features the self-trapping phenomenon when the
coupling strength and/or particle momentum is varied. For the
first time accurate data for energy, effective mass, the structure
of the polaronic cloud, dispersion law, and spectral
function are available
throughout the crossover region. We observed that self-trapping
can not be reduced to hybridization of two states with different
lattice deformation, and that at least three states are involved
in the crossover from light- to heavy-mass regimes.
\end{abstract}
\maketitle

Properties of particles strongly coupled to their environment are
of importance in many fields of physics and are attracting
constant attention given extreme diversity of what may be called a
``particle,'' an ``environment,'' and how they interact with each
other. In most general terms, self-trapping (ST) means a dramatic
transformation of particle properties when system parameters are
slightly changed. 
Landau \cite{landau} showed that the ``trapped''
(T) particle state
with strong lattice deformation around it and the weakly perturbed
``free'' (F) particle state may have the same energy at some critical 
value of the coupling strength, $\alpha_c$.

Of course, the resonance between F and T states is not infinitely
sharp since the matrix element hybridizing them is non-zero, i.e.
ST phenomenon is a crossover, rather than a transition, and all
polaron properties are analytic in $\alpha$---see
Ref.~\onlinecite{Gerlach} for an explicit proof. This theorem
makes the notion of ST rather vague since there is always some
admixture of one state in another. Moreover, it challenges the
adopted opinion that only two, namely F and T, states are in
competition. If there are more than two states within the
energy scale of the hybridization matrix element 
then all of them are mixed and the F-T classification fails. 
In fact, the two-states assumption on which the current theory
is based is not supported by experiments and 
rather complex spectra are usually observed instead
\cite{Matsui,Koda}.

According to the standard criterion \cite{R82,UKKTH86}, ST takes
place if there is a barrier $U_B$ in the adiabatic potential
between the bare-particle and  polaron states.  It occurs, almost
by definition, in the intermediate coupling regime where
perturbation theory is not applicable. Hence, the existence of a
barrier---if the very notion of the adiabatic potential is not ill
defined---and the ST phenomenon can be addressed only by an exact
method, because  in the intermediate coupling regime an analytic
solution is hardly available, and even sophisticated variational
treatments often give misleading results \cite{oops}.

In this Letter, we consider a typical model in which particle
couples to the environment of gapped dispersionless optical
phonons. For this model it is possible to define ST in a mathematically
rigorous way and proceed with its quantitative study. We show how
various particle properties (energy, effective mass, dispersion
law, and the structure of the polaronic cloud) change between
weak- and strong-coupling limits, and provide detailed information
about ST of polarons, which is not based on any approximations.
Besides, we show that there are at least three states involved in
mixing in the critical region and, thus commonly accepted concept
of only F and T states mixing at $\alpha_c$  appears to be
oversimplified. In fact, we are not aware of any other numerical
study testing how accurate are existing treatments of the ST
problem.


The Hamiltonian of the system consists of the free-particle term
(we consider continuum three-dimensional case with dispersion
relation $\varepsilon({\bf k})=k^2/2m$ )
\begin{equation}
H_{\mbox{\scriptsize e}} \, = \, \sum_{\bf k} \, \varepsilon({\bf
k}) \, a^{\dag}_{\bf k} a^{ }_{\bf k} \; , \label{e}
\end{equation}
the Hamiltonian of the phonon bath
\begin{equation}
H_{\mbox{\scriptsize ph}} \, = \,
\sum_{\bf q} \, \omega_q \, b^{\dag}_{\bf q}  b^{ }_{\bf q}
= \omega_0  \sum_{\bf q} \, b^{\dag}_{\bf q}  b^{ }_{\bf q} \; ,
\label{ph}
\end{equation}
and the standard density-displacement interaction \cite{polaron}
\begin{equation}
H_{\mbox{\scriptsize e-ph}} \, = \,
\sum_{{\bf k},{\bf q}} \, V({\bf q}) \,
\left( b^{\dag}_{\bf q} - b^{ }_{-{\bf q}}   \right) \,
a^{\dag}_{{\bf k}-{\bf q}} a^{ }_{\bf k} \;.
\label{e-ph}
\end{equation}
In Eqs.~(\ref{e}-\ref{e-ph}), $a^{ }_{\bf k}$ and $b^{ }_{\bf q}$
are the particle and phonon annihilation operators in momentum
space, correspondingly.

Our study is based on the quantum Monte Carlo simulation of the
polaron Green function in imaginary time at $T=0$ and 
subsequent analytic continuation to the real frequencies
\cite{PS,MPSS,QDST}. The method suggested in 
Refs.\ \cite{PS,MPSS,QDST} is free from approximations and 
systematic errors. It
is particularly suited for the study of ST problem where several
$\delta$-peaks are expected below the spectral continuum in the Lehman
expansion
\begin{equation}
S^{({\bf k})} (\omega) \, = \, \sum_{\nu} \, \delta(\omega -
E_{\nu}({\bf k})) \; \vert \langle \nu \vert a^{\dag}_{\bf k}
\vert \mbox{vac} \rangle \vert^2 \; .
\label{g}
\end{equation}
Here $\{ \vert \nu \rangle \}$ is a complete set of eigenstates of
$H$ in the momentum sector ${\bf k}$, i.e. ~$H \, \vert \nu ({\bf
k}) \rangle = E_{\nu}({\bf k}) \, \vert \nu ({\bf k}) \rangle$.

Separating stable quasiparticle states (labeled by index $i$) from
continuum, we rewrite Eq.~(\ref{g}) as
\begin{equation}
S^{({\bf k})} (\omega) \, = \,\sum_{i} Z_i^{\bf k}(0) \delta
\left( \omega - E^{({\bf k})}_{i}\right) + \int_{\omega_c} d\omega
\,  s^{({\bf k})} (\omega ) \; , \label{gpol}
\end{equation}
where $Z_i^{\bf k}(0)$ and $E^{({\bf k})}_{i}$ are $Z$-factors and
energies of stable states, and the continuum threshold is given by
$\omega_c = E^{({\bf k=0})}_{0} + \omega_0$.  Any state with $E
> \omega_c$ is unstable against single- ($n=1$)  or multi-phonon
($n>1$) emission process $E \to E^{({\bf p})}_{\rm i}+ n
\omega_0$,  where momentum ${\bf p}$ is selected only by the
energy conservation law since phonons are dispersionless.

Speaking rigorously, by self-trapping one understands the
existence of such a region in the parameter space of $H$ where
more than one stable polaron states, differing by the degree of
polarization of the lattice, coexist. This definition implies
three critical points in the coupling constant (keeping other
parameters fixed for simplicity), $\alpha_{c1}({\bf k}) <
\alpha_c({\bf k}) < \alpha_{c2}({\bf k})$. The ST ``transition
point" $\alpha_c$ is understood as the point of avoided crossing
between the two lowest polaron states. At this point the
groundstate of the polaron is a hybrid of states with
substantially different degrees of lattice polarization. The
critical points $\alpha_{c1}$ and $\alpha_{c2}$ correspond to the
appearance and disappearance of the extra stable state(s),
respectively. [By definition, the energy difference $\Delta
E^{({\bf k})} (\alpha)$ between the ground and first stable
excited state, which has its minimum at $\alpha_c ({\bf k}) $,
ought to be less than $\omega_c-E^{({\bf k})}_{0}$.] Critical
couplings introduced above are consistent with previous
considerations \cite{landau,R82,UKKTH86} and have an advantage of
being unambiguous even when the minimal gap $\Delta E^{({\bf k})}
(\alpha_c({\bf k}))$ is not small \cite{aku}.

A typical system that is believed to feature ST is the so-called
Rashba-Pekar model \cite{PekarRa,DyPe} which describes Wannier
exciton in the  $1s$ state interacting with optical vibrations via
electrostatic potential \cite{UKKTH86}
\begin{equation}
V({\bf q}) =\gamma ({\bf q}) \left\{
{1 \over [1+(\xi_e a_Bq)^2 ]^{2} }
 - {1 \over [1+(\xi_h a_Bq)^{2} ]^{2} } 
 \right\}, \label{V}
\end{equation}
\begin{equation}
\gamma({\bf q}) \, = \, i \, \left( 2 \sqrt{2} \alpha \pi \right)^{1/2}  \,
q^{-1} \; .
\label{ga}
\end{equation}
Here $\alpha$ is the standard dimensionless coupling constant,
$a_B$ is the Bohr radius, and $\xi_{e,h}=m_{e,h}/[2(m_e+m_h)]$ is
given in terms of electron ($m_e$) and hole ($m_h$) masses,
respectively.
\begin{figure}
\includegraphics{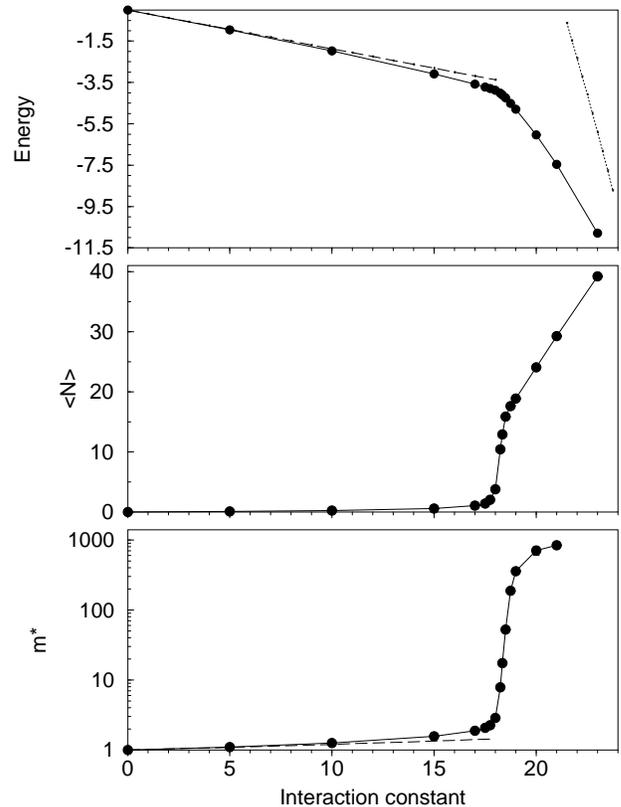} 
\caption{\label{fig:fig1}  The
groundstate energy, average number of phonons, and effective mass
as functions of $\alpha$ (points connected by solid lines).
Relative statistic errors are less than $10^{-3}$ and $10^{-2}$
for the energy and $\langle N \rangle$, respectively. The relative
statistic errors for the mass are of order $10^{-2}$ for
$\alpha<18.5$ and around $5 \times 10^{-2}$ for larger coupling
constants. Dashed lines show results of the perturbation theory
while the dotted line corresponds to the strong-coupling limit. }
\end{figure}
In this Letter, we focus on the parameters corresponding to the
curve number 2 of Ref.~\cite{PekarRa},
which is believed to describe ST in the strong-coupling regime.
Setting the total bare mass of the exciton $m=m_e+m_h$, phonon
frequency $\omega_0$, electric charge, and Plank constant to
unity, one finds that $m_e=0.065$.  The Bohr radius can be used to
change the degree of adiabaticity in the model. Below  we shall
thoroughly consider an ``almost adiabatic'' case (i) with
$U_B/\omega_0=2$ (which is realised for $a_B=0.467$) and outline
some peculiar features of the ``nonadiabatic'' situation (ii) with
$U_B/\omega_0=0.5$ ($a_B=0.934$). The critical coupling constants,
determined within the approach of Ref.~\cite{PekarRa}, are then
$\alpha_c^{\mbox{\scriptsize ad}} \approx 14.3$ and
$\alpha_c^{\mbox{\scriptsize ad}} \approx 7.2$, respectively.

In Fig.~1 we show how the groundstate properties (${\bf k}=0$)
depend on the coupling strength. The groundstate energy, the
effective mass $m^*$, and the average number of phonons in the
polaronic cloud
\begin{equation}
\langle N \rangle = \langle {\bf k}=0 \vert \, \sum_{\bf q}\,
b^{\dag}_{\bf q}  b^{ }_{\bf q} \, \vert {\bf k}=0 \rangle,
\label{n_av}
\end{equation}
clearly indicate drastic changes around $\alpha_c \approx 18.35$.
At this point the energy derivative changes very fast, and both
$\langle N \rangle$ and $m^*$ ungergo step-wise increase visible
even on the logarithmic plot for $m^*$. In a narrow region between
$\alpha=17.5$ and $\alpha =19$ the effective mass increases by two
orders of magnitude. The remarkable fact is that at $\alpha_c$ the
strong-coupling approach is still far from being accurate (dotted
line \cite{DyPe} in the upper panel of Fig.~\ref{fig:fig1}).
Besides, the adiabatic critical constant
$\alpha_c^{\mbox{\scriptsize ad}} \approx 21$ \cite{DyPe} differs
significantly from our value $\alpha_c \approx 18.35$. For the
``nonadibatic'' case (ii) the behavior of the groundstate
properties is qualitatively the same, but quantitative deviations
from the strong-coupling limit are larger.

\begin{figure}
\includegraphics{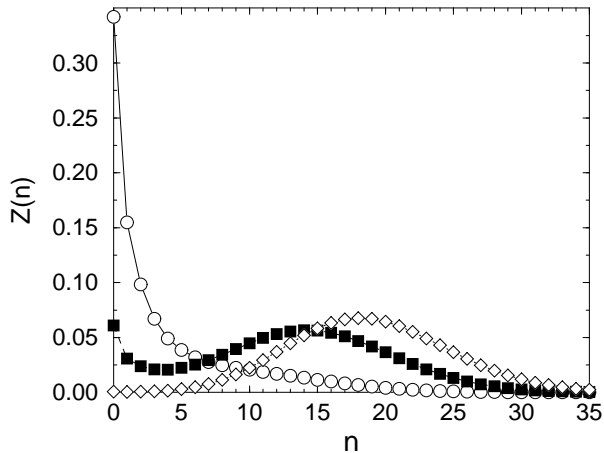}
\caption{\label{fig:fig2}
Partial weights of n-phonon states in the polaron ground
state (${\bf k}=0$) at $\alpha=18$ (circles),
$\alpha=18.35$ (squares), and
$\alpha=19$ (dimonds).
Statistic errorbars of order $3 \times 10^{-3}$ are less than
the symbol size.}
\end{figure}

Next, we study how the phonon cloud evolves throughout the ST
critical region. Partial $n$-phonon contributions to the polaron
ground state $Z_0^{\bf k}(n)$ are the probabilities of finding
exactly $n$ phonons in the cloud, and the average number of
phonons introduced earlier, is just $\langle N \rangle = \sum_n n
Z(n)$. Figure \ref{fig:fig2} shows $Z_0^{{\bf k}=0}(n)$
distributions at $\alpha=18$ (below the crossover region),
$\alpha=\alpha_c=18.35$, and  $\alpha=19$ (trapped state). We see
that the distribution at $\alpha_c$ has two peaks and is half-way
between the two limiting cases. However, in the ``nonadiabtic''
case (ii) the structure with two maxima in $Z(n)$ is missing.
Therefore, the peculiar behavior presented in Fig.~\ref{fig:fig1}
is a general feature of the ST phenomenon whereas the two-peak
structure of the phonon distribution is specific for the adiabatic
limit.
\begin{figure}
\includegraphics{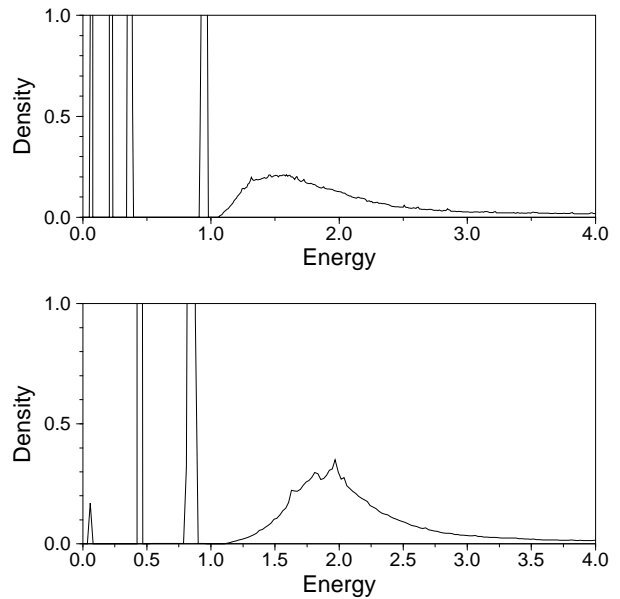}
\caption{\label{fig:fig3}
The Lehman spectral function $S^{({\bf k}=0)}$
at coupling constants
$\alpha=18.35$ (upper panel) and $\alpha=18.75$ (lower panel).}
\end{figure}

\begin{figure}
\includegraphics{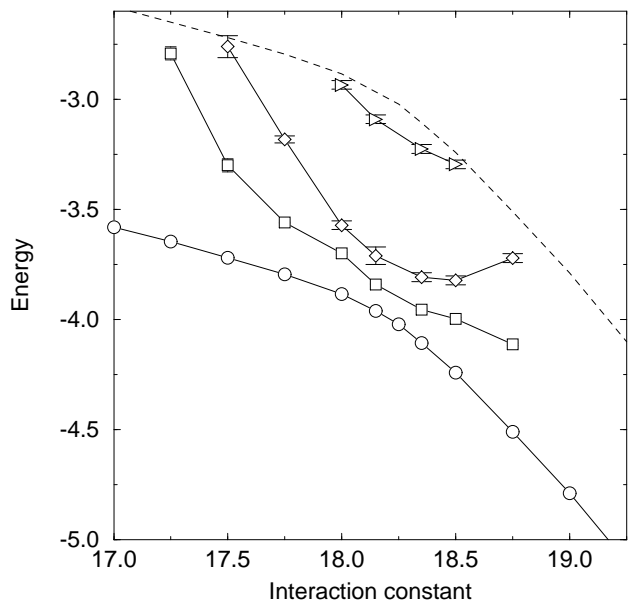} 
\caption{\label{fig:fig4}
Energies of
the ground (circles) and excited stable states (squares, diamonds,
and triangles) {\it vs}
 interaction constant. The
dashed line is the threshold of incoherent continuum. Typical
errorbars for the first, second, and third excited states are
$10^{-2}$, $3 \times 10^{-2}$, and $4 \times 10^{-2}$, respectively.}
\end{figure}

The spectral function around the critical point \cite{cond-mat}
reveals up to three stable excited states below the continuum
threshold (see examples of the Lehman function in
Fig.~\ref{fig:fig3}). We observe in Fig.~\ref{fig:fig4} that three
polaronic states (in the energy range comparable with the
hybridization strength) participate in the ST crossover. We
underline that all three  states have large $Z$-factors ($>0.1$).
Therefore, for the given set of parameters more than two states
are mixed at the crossover point and the standard picture of F-T
hybridization at the tip of the ST crossover fails. 
One can speculate that extra stable states in the gap, which 
standard theory puts into the spectral continuum, are due to 
excited levels of highly nonlinear ST potential in the 
resonating region. However, this interpretation is essentially 
qualitative since the concept of adiabatic potential breaks down 
in the crossover region.
\begin{figure}
\includegraphics{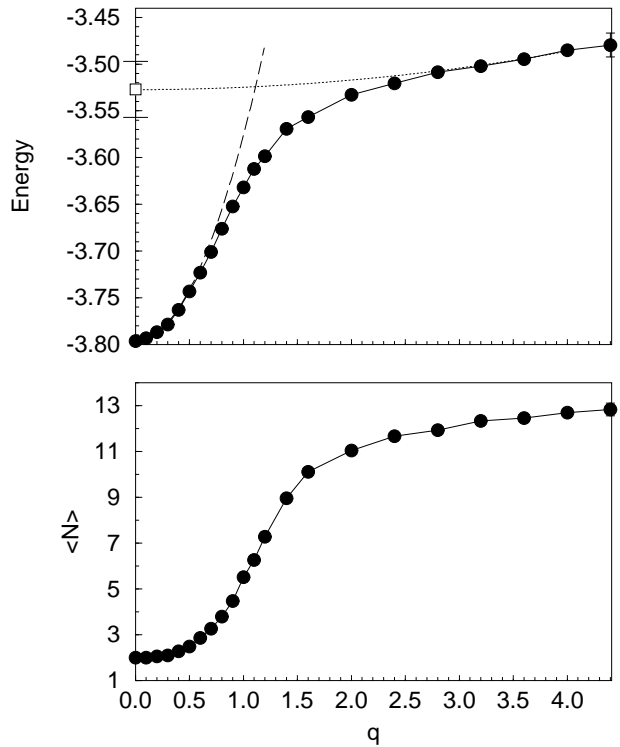} 
\caption{\label{fig:fig5}
The
wave-vector dependence of energy and average number of phonons for
$\alpha=17.75$ (circles connected by solid lines). The errorbars
are $3 \times 10^{-3}$ and $10^{-2}$ for energy and $\langle N
\rangle$, respectively. The dashed curve is the effective mass
approximation $E^{({\bf k})}= E_0 + {\bf k}^2/2m^*$ with
$E_0(\alpha=17.75)=-3.7946$ and $m^*(\alpha=17.75)=2.258$ obtained
from direct Monte Carlo estimators. The dotted curve is the
parabolic dispersion law fitted to the last four points in the
energy plot with parameters $E_1(\alpha=17.75)=-3.5273$ and
$m^*_1=195$. The open rectangle is the energy obtained from
spectral analysis for the first excited state.}
\end{figure}

So far we have considered ST crossover at zero momentum, i.e., for
the ground state. However, same considerations apply to finite
momentum states, as long as $k<k_c$ where $k_c$  is defined as the
point where $E_0^{({\bf k})}=\omega_c$ and the polaron spectrum
has an end point \cite{Larsen,PS}. For $\alpha < \alpha_c$ the
ground state is characterized by small lattice distortion and
light effective mass and thus the dispersion law associated with
this state rises steep with ${\bf k}$. On another hand, the
dispersion curve of the excited heavy-mass state is nearly flat
which means that light- and heavy-branches have to intersect at
some wave vector. In Fig.~\ref{fig:fig5} we plot the groundstate
dispersion law and the average number of phonons in the ground
state for $\alpha = 17.75$ which we interpret as the level
crossing picture (see the figure caption). Notice the agreement
between the spectral analysis at ${\bf k}=0$ and the parabolic fit
of the heavy mass branch.

By all accounts, the existence of more than one stable polaron
state is a highly nontrivial qualitative property of the model
whether the hybridization gap between these states is small, or
not. Moreover, the ST crossover is not necessarily limited to
hybridization of only two states. 
Recent studies of the Holstein polaron in 1D
strongly support the universality of the latter statement
since more than two stable states were found there 
as well \cite{Macri}.

We found that the dependence of
$E_0$ and $\langle N \rangle$ (but not the structure of the cloud)
on coupling can be used as an indirect indication of the self
trapping phenomenon with a well-defined ``transition point" (the
point of minimal gap between the ground and first stable excited
state). It was considered previously as a theorem, that ST may
occur only in dimensions $d>2$ \cite{R82,UKKTH86}; however, the
definition of what has to be counted as a ST  transition was not
given in quantitative terms. We believe that it is more
appropriate to use less ``radical", but unambiguous, definition
adopted in our paper. Recently, the second stable polaron state
was found to exist for the Holstein polaron in a one-dimensional
lattice \cite{BoTru99} and in infinite dimension approximation
\cite{CiuPas95}.

This work was supported by the National Science Foundation under Grant
DMR-0071767 and RFBR No 01-02-16508.


\begin{thebibliography}{99}
\bibitem{oops} Self-trapping was predicted to occur in the Fr\"{o}hlich
               polaron model by various papers, but recently it has been proved
               to be an artifact of the variational scheme \cite{MPSS}.
\bibitem{aku} With non-zero coupling
to acoustic modes the excited polaron states have finite
lifetimes. However, when the energy gap between the ground and
excited states gets small, the life-time diverges due to vanishing
probability of emitting a long-wavelength phonon.
\bibitem{cond-mat} The specific features of the spectral analysis
in the present case will be presented in the subsequent paper.
\end{thebibliography}
\end{document}